# Efficient and ultra-stable perovskite light-emitting diodes

Bingbing Guo[1], Runchen Lai[1], Sijie Jiang[2,3], Yaxiao Lian[1], Zhixiang Ren[1], Puyang Li[2,3], Xuhui Cao[1], Shiyu Xing[1], Yaxin Wang[1], Weiwei Li[4,5], Chen Zou[1], Mengyu Chen[2,3], Cheng Li[2,3], Baodan Zhao[1]*, Dawei Di[1]*

1. State Key Laboratory of Modern Optical Instrumentation, College of Optical Science and Engineering; International Research Center for Advanced Photonics, Zhejiang University, Hangzhou, 310027, China
2. School of Electronic Science and Engineering, Xiamen University, Xiamen, 361005, China
3. Future Display Institute of Xiamen, Xiamen, 361005, China
4. MIIT Key Laboratory of Aerospace Information Materials and Physics, College of Physics, Nanjing University of Aeronautics and Astronautics, Nanjing, 211106, China
5. Department of Materials Science and Metallurgy, University of Cambridge, Cambridge, CB3 0FS, United Kingdom

*Corresponding author E-mail: daweidi@zju.edu.cn (D.D.); baodanzhao@zju.edu.cn (B.Z.)

**Perovskite light-emitting diodes (PeLEDs) have emerged as a strong contender for next-generation display and information technologies. However, similar to perovskite solar cells, the poor operational stability remains the main obstacle toward commercial applications. Here we demonstrate ultra-stable and efficient PeLEDs with extraordinary operational lifetimes ($T_{50}$) of $1.0\times10^4$ h, $2.8\times10^4$ h, $5.4\times10^5$ h, and $1.9\times10^6$ h at initial radiance (or current densities) of 3.7 W sr$^{-1}$ m$^{-2}$ (~5 mA cm$^{-2}$), 2.1 W sr$^{-1}$ m$^{-2}$ (~3.2 mA cm$^{-2}$), 0.42 W sr$^{-1}$ m$^{-2}$ (~1.1 mA cm$^{-2}$), and 0.21 W sr$^{-1}$ m$^{-2}$ (~0.7 mA cm$^{-2}$) respectively, and external quantum efficiencies of up to 22.8%. Key to this breakthrough is the introduction of a dipolar molecular stabilizer, which serves two critical roles simultaneously. First, it prevents the detrimental transformation and decomposition of the α-phase $FAPbI_3$ perovskite, by inhibiting the formation of lead and iodide intermediates. Secondly, hysteresis-free device operation and microscopic luminescence imaging experiments reveal substantially suppressed ion migration in the emissive perovskite. The record-long PeLED lifespans are encouraging, as they now satisfy the stability requirement for commercial organic LEDs (OLEDs). These results remove the critical concern that halide perovskite devices may be intrinsically unstable, paving the path toward industrial applications.**

Light-emitting diodes (LEDs) based on perovskite semiconductors have shown great promises as next-generation light sources, as they combine the advantages of high efficiency and spectral tunability at low processing costs[1-20]. Since the report of room-temperature electroluminescence (EL) from halide perovskites in 2014[1], the development in the field has been fast. The external quantum efficiencies (EQEs) of PeLEDs exceeded the 20% milestone in 2018[4-7], followed by recent improvements of device EQEs to over 28%[21].

Despite the unprecedented pace of development leading to the high device efficiencies, the poor operational lifetimes of PeLEDs remain to be a critical challenge. Typical $T_{50}$ lifetimes (time required for the EL intensity to reach 50% of its initial value) for state-of-the-art PeLEDs under continuous operation are on the order of 10-100 h[4-9, 14-16, 20-22]. Recent reports have shown that it is possible to achieve $T_{50}$ of 682 h at an initial radiance of 17 W sr$^{-1}$ m$^{-2}$ using a dicarboxylic acid additive[23]. Practical applications demand longer operational lifetimes ($T_{50}$ = 1000-10000 h or higher) at high efficiencies (EQE > 10%) for a wide range of radiance.

Investigations into the mechanisms of PeLED degradation are still at an early stage[24-26]. In comparison to III-V and organic semiconductors, metal halide perovskites present some additional degradation mechanisms in device operation. The migration of ionic species under electric field, and the instability of the perovskite crystal structures are some of the key issues affecting the stability of perovskite solar cells[27-32] and LEDs[22-26]. Resolving these problems to realize long device lifetimes while retaining high EL efficiencies for ideal LED operation remains a significant challenge.

In this work, we report high-performance near-infrared (NIR) PeLEDs with ultralong intrinsic operational lifetimes. The PeLEDs were prepared with a structure of glass/indium tin oxide (ITO)/polyethylenimine ethoxylated (PEIE)-modified zinc oxide (ZnO)/perovskite/poly(9,9-dioctyl-fluorene-co-n-(4-butylphenyl) diphenylamine) (TFB)/molybdenum oxide (MoO$_x$)/gold (Au) (Fig. 1a and Extended Data Fig. 1). The precursor solution for the perovskite emissive layer was prepared by dissolving formamidinium iodide (FAI), lead iodide (PbI$_2$) and sulfobetaine 10 (SFB10) (a dipolar molecule) at a molar ratio of 2:1:x (x = 0-0.3) in N, N-dimethylformamide (DMF). The molecular structure of SFB10 is shown in Fig. 1b. The molar fraction of the SFB10 in the precursor solution was optimized to be x=0.15 (Extended Data Fig 2). In this paper, samples without SFB10 are denoted 'control' samples. PeLEDs with and without SFB10 show nearly identical EL spectra peaked at 803 nm (Fig. 1c, inset & Extended Data Fig. 2a), indicating negligible effects of SFB10 on the perovskite bandgap.

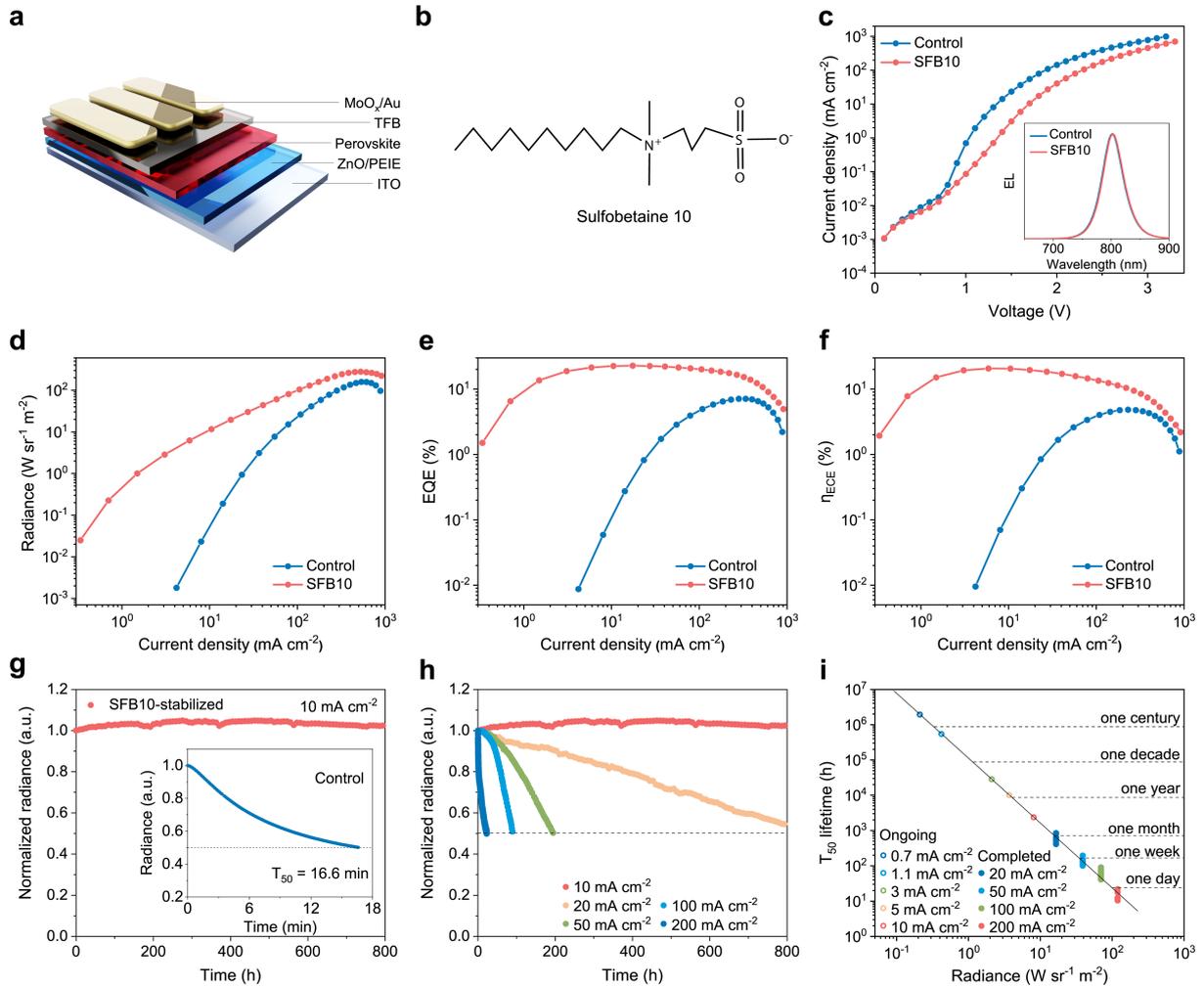

**Fig. 1 | The structure and performance of PeLEDs. a**, The device structure of PeLEDs. **b**, The molecular structure of SFB10. **c**, Current density-voltage curves. Inset shows the EL spectra of control and SFB10 devices. **d**, Radiance-current density characteristics. **e**, EQE-current density characteristics. **f**, ECE-current density curves. **g**, Operational stability tests for SFB10 devices under 10 mA cm$^{-2}$. Inset shows the stability performance for control devices. **h**, Accelerated aging tests for SFB10-stabilized PeLEDs at different current densities. The measurements were performed in a N$_2$ glovebox. **i**, The T$_{50}$ lifetimes as a function of initial radiance (R$_0$), the solid line is the fitting of T$_{50}$ data to equation R$_0^n$ × T$_{50}$ = constant, where n is the acceleration factor (n~1.84). The data points marked by solid dots are from completed T$_{50}$ measurements (35 data points in total). The open circles are the T$_{50}$ lifetimes for the ongoing measurements at medium and low current densities, which are expected to finish after longer times.

The performance characteristics of the PeLEDs with and without SFB10 are shown in Fig. 1c-f. The SFB10-based PeLEDs show peak EQEs of up to 22.8% (record-high for NIR PeLEDs) and maximum radiance of 278.9 W sr$^{-1}$ (Fig. 1d, e). These are substantially higher than that achieved for the control devices (peak EQE= 7.1%, maximum radiance = 158.3 W sr$^{-1}$ m$^{-2}$). We note that the EQE values for the SFB10-based PeLEDs are high (10% - 22.8%) across a wide range of current densities (~1 to ~500 mA cm$^{-2}$) or radiance values (~0.52 to ~278 W sr$^{-1}$ m$^{-2}$), comparing favourably with state-of-the-art NIR PeLEDs[5,6,8,22,23]. The PeLEDs with excellent low-, medium- and high-current-density performance show record-high energy-conversion efficiencies ($\eta_{ECE}$, the conversion efficiency from electrical power to light) of up to 20.7% (Fig. 1f). Compared to the control devices, SFB10-based PeLEDs show considerably improved average EQEs (Extended Data Fig. 2e). The angular emission intensity of our perovskite LEDs follows a Lambertian profile (Extended Data Fig. 2f).

The operational stability tests of the PeLEDs were carried out in a $N_2$-filled glovebox. The EL intensity from the SFB10-stabilized PeLEDs driven under a constant current density of 10 mA $cm^{-2}$ (initial radiance $R_0$ = 8.1 W $sr^{-1}$ $m^{-2}$) showed no observable degradation over 800 h of continuous operation (the measurements are ongoing) (Fig. 1g). In contrast, the EL intensity of the control devices reduced rapidly with a $T_{50}$ of 16.6 min under the same test conditions. To determine the operational lifetimes of the SFB10-stabilized PeLEDs, accelerated aging tests were performed at a range of higher current densities (Fig. 1h). Maximum $T_{50}$ lifetimes at current densities (or initial radiance) of 200 mA $cm^{-2}$ (~119 W $sr^{-1}$ $m^{-2}$), 100 mA $cm^{-2}$ (~70 W $sr^{-1}$ $m^{-2}$), 50 mA $cm^{-2}$ (~39 W $sr^{-1}$ $m^{-2}$) and 20 mA $cm^{-2}$ (~17 W $sr^{-1}$ $m^{-2}$) are 22.4 h, 120.3 h, 195.3 h and 877.1 h, respectively (Extended Data Fig. 3, 4a). The average $T_{50}$ lifetimes are summarised in Extended Data Table 1. The radiance-current density relationship was determined from the averaged results of 10 representative devices (Extended Data Fig. 3). The EL spectrum remained unchanged after 22.4 h of continuous operation under a high current density of 200 mA $cm^{-2}$ (Extended Data Fig. 4b). An empirical scaling law widely used for describing LED degradation, $R_0^n \times T_{50}$ = constant (where n is the acceleration factor)[33-34], is used in modelling the degradation behaviour of our PeLEDs. The statistical average of the lifetime data at a range of radiance values can be fitted satisfactorily with this model (Fig. 1i). While LED aging tests for lower current and radiance settings are still ongoing, $T_{50}$ lifetimes of $1.0 \times 10^4$ h, $2.8 \times 10^4$ h, $5.4 \times 10^5$ h and $1.9 \times 10^6$ h can be estimated for driving conditions of 5 mA $cm^{-2}$ (~3.7 W $sr^{-1}$ $m^{-2}$), 3.2 mA $cm^{-2}$ (~2.1 W $sr^{-1}$ $m^{-2}$), 1.1 mA $cm^{-2}$ (~0.42 W $sr^{-1}$ $m^{-2}$) and 0.7 mA $cm^{-2}$ (~ 0.21 W $sr^{-1}$ $m^{-2}$), respectively (Fig. 1i & Extended Data Table 1). These results represent a record for the operational stability of PeLEDs.

To understand the origins of the high operational stability, we investigated the properties of the perovskite emissive layers. To examine the stability of the perovskite materials under elevated temperatures, which is expected to occur during device operation, we placed the perovskite samples on a hotplate set to 100 °C in a $N_2$ glovebox. For the control samples (Fig. 2a), after 90-min annealing we observed the emergence of peaks at around 12.6°, corresponding to the (001) diffraction peaks of $PbI_2$. The diffraction peaks grew in intensity over time, indicating the continued generation of $PbI_2$ in the control samples. In contrast, for SFB10-stabilized perovskite samples (Fig. 2b), no $PbI_2$ diffraction peaks were observed after 360 min of continuous heating. The XRD diffraction peaks at 14.0° corresponding to the (100) planes of the α-phase $FAPbI_3$ perovskite (in agreement with Extended Data Fig. 1b)[31, 32], maintained or increased in intensity during the temperature treatment, indicating excellent thermal stability under elevated temperatures.

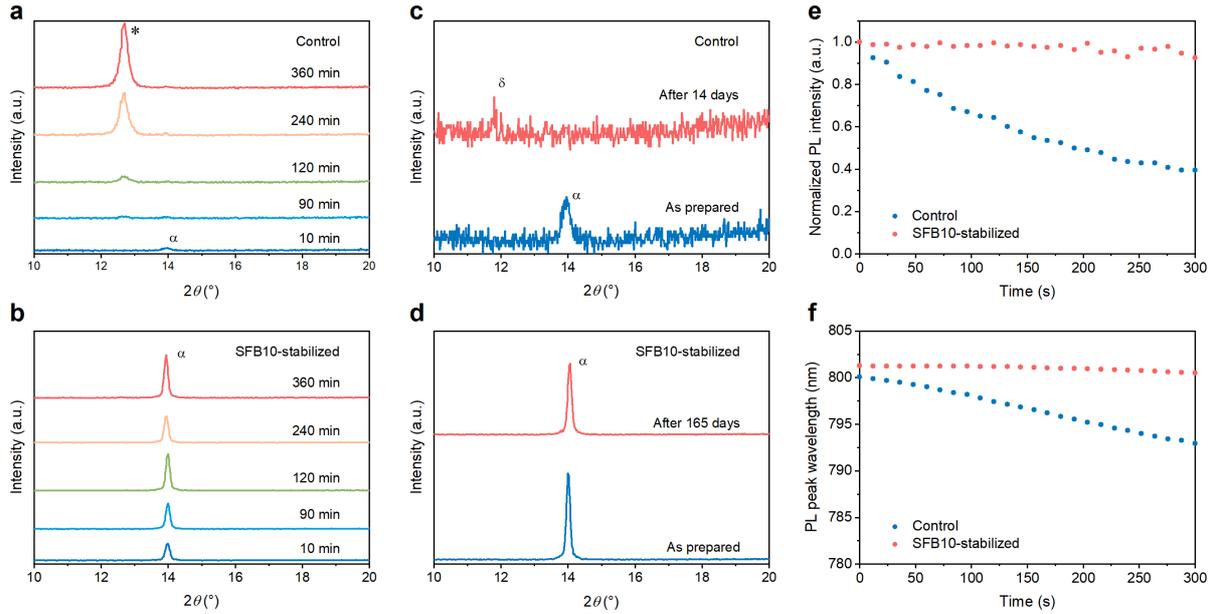

**Fig. 2 | Structural and PL stability of perovskite samples. a**, XRD patterns of control samples after different durations of annealing at 100 °C. * and α denote the diffraction peaks corresponding to PbI$_2$ and α-FAPbI$_3$, respectively. **b**, XRD patterns of SFB10-stabilized films after different durations of annealing at 100 °C. **c**, XRD patterns of the control samples exposed in air (as prepared and after 14 days). **d**, XRD patterns of the SFB10-stabilized samples exposed in air (as-prepared and after 165 days). **e**, PL intensity measurements for glass/ITO/ZnO/PEIE/perovskite samples under 400-nm pulsed laser excitation (~80 μJ cm$^{-2}$, 50 kHz) in air. **f**, PL peak wavelength measurements for glass/ITO/ZnO/PEIE/perovskite samples under 400-nm pulsed laser excitation (~80 μJ cm$^{-2}$, 50 kHz) in air.

The phase stability of the perovskite films was found to be enhanced by SFB10. For the control samples, the diffraction peak of δ-phase FAPbI$_3$ (11.8°) appeared after 14-day storage in air (Fig. 2c), accompanied by the disappearance of peaks related to the α-phase. In contrast, the SFB10-stabilized perovskite samples showed strong diffraction peaks at 14.0° (associated with the α-phase of FAPbI$_3$) without the emergence of any additional peaks (Fig. 2d) after 165 days of storage in air, indicating excellent phase stability. Further, we examined the stability of the samples under intense pulsed laser excitations (80 μJ cm$^{-2}$ at 50 kHz) in air. For perovskite emissive layers deposited on glass/ITO/ZnO/PEIE substrates, SFB10 stabilizer was found to significantly improve the PL stability of the perovskite samples (Fig. 2e and f).

While showing different surface morphologies (Extended Data Fig. 5 & 6), the absorption spectra of the control and SFB10-stabilized films exhibit similar features (Fig. 3a). The transient absorption spectra of the SFB10-stabilized films are shown in Extended Data Fig. 7. A ground-state bleach at 790 nm was observed, consistent with the band-to-band absorption of the FAPbI$_3$ perovskite. The samples with SFB10-stabilized show stronger PL and a longer effective PL lifetime (the time required for the PL intensity to reach 1/e of the initial intensity) of 728 ns (versus 265 ns for the control samples) (Fig. 3b), consistent with reduced density of traps for the SFB10-stabilized perovskite. The current-voltage (J-V) characteristics of electron-only devices based on perovskite layers with and without SFB10 stabilizer are shown in Extended Data Fig. 8. The trap-filled limit voltage ($V_{TFL}$) is reduced to 1.19 V for the SFB10-stabilized samples (from 1.47 V for the control samples), consistent with the reduction of trap densities in these samples.

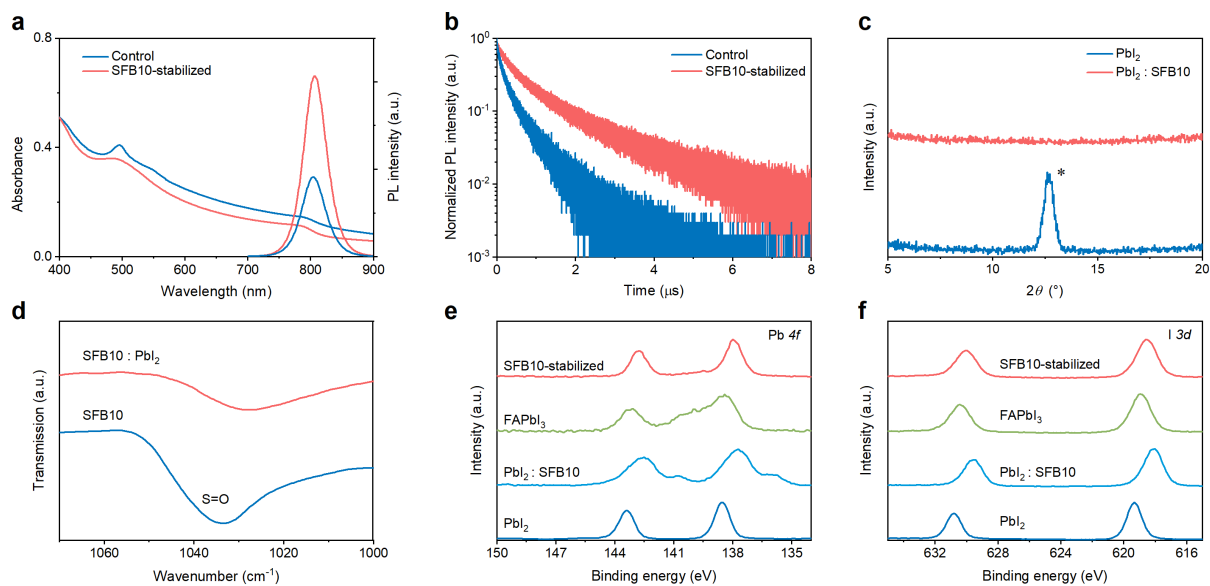

**Fig. 3. Additional characterizations for SFB10-treated samples**. **a**, Absorbance and PL spectra of the perovskite films. **b**, PL decay curves of perovskite films (excitation fluence: 20 nJ cm$^{-2}$). **c**, XRD patterns of pristine PbI$_2$ and SFB10-treated PbI$_2$ samples (the molar ratio of SFB10 to PbI$_2$ was 1:1). * denotes the diffraction peak corresponding to PbI$_2$. **d**, ATR-FTIR spectroscopy of the SFB10 sample and the SFB10-treated PbI$_2$ sample (the molar ratio of SFB10 to PbI$_2$ was 1:1). **e**, XPS spectra of the Pb *4f* peaks for SFB10-stabilized FAPbI$_3$, FAPbI$_3$, SFB10-treated PbI$_2$ and PbI$_2$. **f**, XPS spectra of the I *3d* peaks for SFB10-stabilized FAPbI$_3$, FAPbI$_3$, SFB10-treated PbI$_2$ and PbI$_2$.

Chemical interactions between the SFB10 stabilizer and perovskite precursors were studied to gain further insight into the roles of the stabilizer. The absence of the PbI$_2$ XRD peak (at 12.6°) for samples prepared from mixed PbI$_2$:SFB10 solution suggests the inhibition of PbI$_2$ crystallite formation in the presence of SFB10 (Fig. 3c & Extended Data Fig. 10). From the attenuated total reflectance-Fourier transform infrared spectroscopy (ATR-FTIR) measurements (Fig. 3d), we observed that the S=O stretching vibration peak at 1034 cm$^{-1}$ shifted to 1027 cm$^{-1}$. This could be attributed to the lone electron donation from O of S=O to the empty orbitals of Pb$^{2+}$. X-ray photoelectron spectroscopy (XPS) analysis was conducted to further study the chemical interactions. The observation of the S *2p* peaks confirms the presence of SFB10 in the perovskite samples (Extended Data Fig. 9a, consistent with Extended Data Fig. 1c). Fig. 3e and 3f show the XPS spectra of Pb *4f* and I *3d* peaks of SFB10-stabilized FAPbI$_3$, pristine FAPbI$_3$, SFB10-treated PbI$_2$, and pristine PbI$_2$ samples on glass/ITO/ZnO/PEIE substrates. For samples with SFB10, redshifted spectra were observed, suggesting the chemical interactions between Pb$^{2+}$ and S=O. Owing to the coordination between S=O and Pb$^{2+}$, slower crystallization rates were observed for samples with SFB10 (Extended Data Fig. 10), partly contributing to the improved crystalline qualities. The ratio of S:Pb on the SFB10-stabilized FAPbI$_3$ sample surfaces was two times higher than the molar ratio of corresponding precursors, indicating the majority of SFB10 was present at the surfaces of the perovskite grains.

Forward and reverse current-voltage (J-V) scans were carried out to investigate the hysteretic behaviour of the perovskite LEDs (Fig. 4a, c). Hysteresis was observed in the first scan cycle for the control PeLEDs, and the hysteretic behaviour became more pronounced with more scan cycles. In contrast, during 30 scan cycles no observable hysteresis was recorded for the PeLEDs stabilized with SFB10. We speculate that the SFB10-stabilized devices are less affected by ion migration, as current-voltage hysteresis is known to be related to the ionic transport in perovskite devices[24,26,27,35-37]. To verify this hypothesis, we carried out microscopic PL imaging[24,29,37] of the perovskite films between two in-plane Au electrodes (spacing = 100 μm), with a constant bias (3.0 V) applied across the electrodes (Fig. 4c, d). For the control sample, a significant degradation of PL intensity occurred near the dark spots and quickly expanded across the perovskite sample over the course of the measurements (Fig. 4c). Such PL degradation behavior for perovskite samples under external bias is evidence for ion migration[24,29,37]. In contrast, for the SFB10-stabilized sample, no significant reduction of PL intensity was observed during the measurements (Fig. 4d). These observations confirm that ion migration is effectively suppressed in the SFB10-stabilized samples, in agreement with the reduced hysteresis of PeLEDs[24].

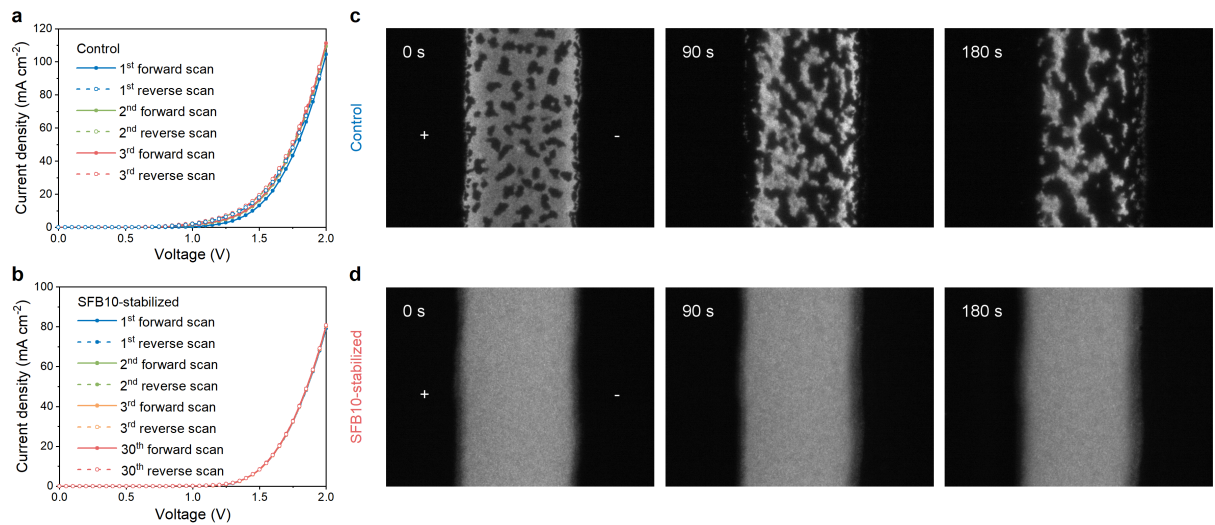

**Fig. 4 | Current-voltage scans of PeLEDs and microscopic PL imaging of perovskite samples. a**, Current density-voltage curves of 3 forward and reverse scan cycles with a scan rate of 0.05 V s$^{-1}$ for the control PeLEDs. **b**, Current density-voltage curves of 30 forward and reverse scan cycles with a scan rate of 0.05 V s$^{-1}$ for the SFB10-stabilized PeLEDs. **c**, Time-dependent PL images of a control sample under an external electric field (~3×10$^4$ V m$^{-1}$) using wide-field PL imaging microscopy. **d**, Time-dependent PL images of a SFB10-stabilized sample under an external electric field (~3×10$^4$ V m$^{-1}$) using wide-field PL imaging microscopy. The polarity of the external electric field is marked by the "+" and "-" symbols. The perovskite samples were excited by a 440-nm c.w. laser (~35 mW cm$^{-2}$). The channel length between the electrodes is ~100 μm.

In summary, we have developed efficient (peak EQE = 22.8%) and ultra-stable PeLEDs with extraordinary $T_{50}$ lifetimes of 1.0×10$^4$ h, 2.8×10$^4$ h, 5.4×10$^5$ h, and 1.9×10$^6$ h at initial radiance (or current densities) of 3.7 W sr$^{-1}$ m$^{-2}$ (~5 mA cm$^{-2}$), 2.1 W sr$^{-1}$ m$^{-2}$ (~3.2 mA cm$^{-2}$), 0.42 W sr$^{-1}$ m$^{-2}$ (~1.1 mA cm$^{-2}$), and 0.21 W sr$^{-1}$ m$^{-2}$ (~0.7 mA cm$^{-2}$), respectively. Key to this breakthrough is the use of a dipolar stabilizer, SFB10, in the preparation of the perovskite emissive films. Beside serving as an agent for trap passivation for efficient LED operation (EQEs of 10% - 22.8%) for a wide range of current densities (~1 to ~500 mA cm$^{-2}$), we have shown that the stabilizer serves two functions. First, it prevents the formation of lead iodide intermediates, which accompany the detrimental phase transformation and decomposition of the α-phase FAPbI$_3$ perovskite. Secondly, we find that the SFB10-stabilized devices show negligible hysteresis and suppressed ion migration. The EL lifetime results satisfy the stability requirement (T$_{50}$ > 10000 h at R$_0$ ~ 0.21-2.1 W sr$^{-1}$ m$^{-2}$) for commercial organic LEDs (OLEDs)[33,38,39], eliminating the critical concern that halide perovskite devices might be intrinsically unstable. The ultralong device lifespans showcase the strong potential of next-generation light-emitting technologies based on perovskite semiconductors.

**Acknowledgements**
This work was supported by the National Key R&D Program of China (grant no. 2018YFB2200401), the National Natural Science Foundation of China (NSFC) (61975180, 62005243, 62005230, 61974126, 51902273, 52102177), Kun-Peng Programme of Zhejiang Province (D.D.), Natural Science Foundation of Zhejiang Province (LR21F050003), Natural Science Foundation of Jiangsu Province (BK20210313), Top-notch Academic Programs Project of Jiangsu Higher Education Institutions (TAPP) & Jiangsu Specially-Appointed Professor Program (W.-W.L.), Fundamental Research Funds for the Central Universities (2019QNA5005, 2020QNA5002, 20720200086, 20720210088), and Zhejiang University Education Foundation Global Partnership Fund. We acknowledge the technical support from the Core Facilities, State Key Laboratory of Modern Optical Instrumentation, Zhejiang University.

**Author contributions**
B.G. planned the experiments under the guidance of D.D. and B.Z. B.G. designed and fabricated the highly stable and efficient PeLEDs, carried out device characterization and data analyses. B.G. and R.L. performed the TCSPC measurements. R.L. performed the TA experiments. S.J. and P.L. conducted the microscopic luminescence imaging experiments under the supervision of C.L. Z.R. assisted with the preparation of substrates for device stability tests. X.C. and B.G. performed the angular emission profile measurements. B.G., B.Z. and D.D. wrote the initial manuscript, with useful inputs from Y.L. All authors contributed to the work and commented on the paper.

**Competing interests**
D.D., B.G., Y.L. and B.Z. are inventors on CN patent application: 202111447766.0.

**Corresponding authors**
Correspondence to: Dawei Di (daweidi@zju.edu.cn), or Baodan Zhao (baodanzhao@zju.edu.cn).

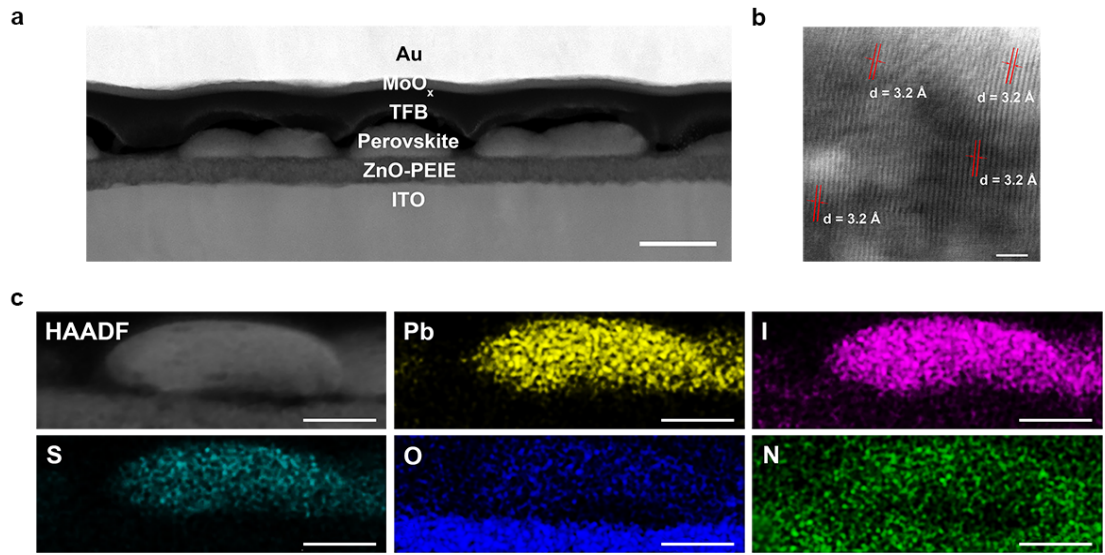

**Extended Data Fig. 1 | HAADF-STEM images and EDS composite mapping of SFB10-stabilized devices. a**, Cross-sectional HAADF-STEM image of a SFB10-stabilized device. Scale bar: 50 nm. **b**, Cross-sectional HAADF-STEM image of a SFB10-stabilized sample. Scale bar: 2 nm. The lattice spacings (10 fringes) show agreement with the cubic α-phase of FAPbI$_3$. **c**, EDS elemental maps of the SFB10-stabilized perovskite device. Scale bar: 50 nm.

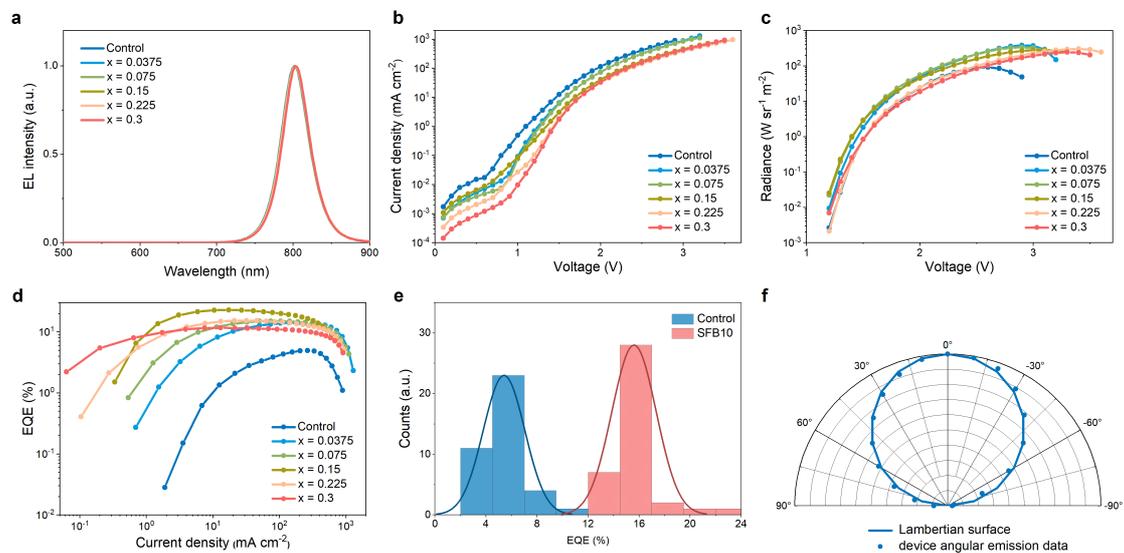

**Extended Data Fig. 2 | Further performance data of control and SFB10-based PeLEDs. a**, EL spectra of PeLEDs with different SFB10 content. **b**, Current density-voltage characteristics. **c**, Radiance-voltage characteristics. **d**, EQE-current density characteristics. **e**, EQE histograms for control and SFB10-based PeLEDs. **f**, Angular emission profile of SFB10-based PeLEDs.

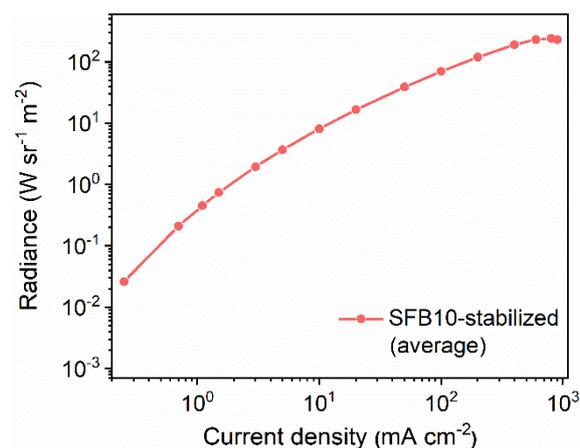

**Extended Data Fig. 3 | The radiance-current density relationship determined from the averaged results of 10 representative devices.**

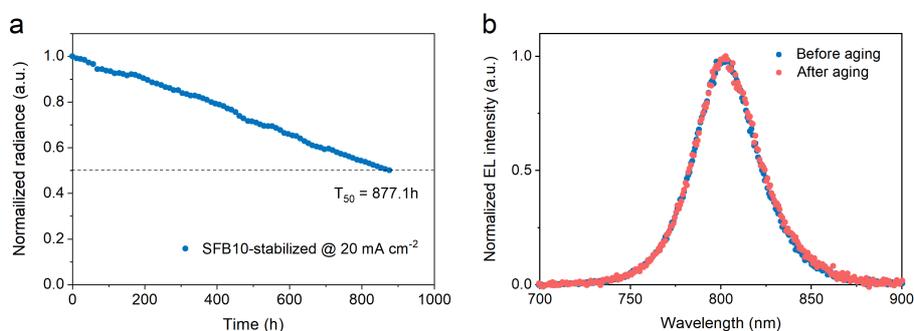

**Extended Data Fig. 4 | Additional operational and spectral stability measurements of SFB10-stabilized PeLEDs. a**, Aging tests of SFB10-stabilized PeLEDs driven at 20 mA cm$^{-2}$. **b**, EL spectra for SFB10-stabilized PeLEDs before and after aging tests at a current density of 200 mA cm$^{-2}$ for 22.4 h.

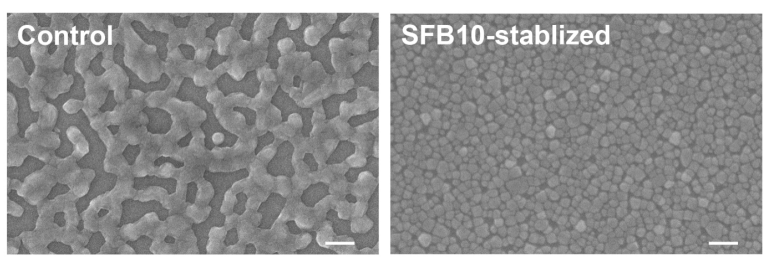

**Extended Data Fig. 5 | SEM images of perovskite film of control and SFB10-stabilized samples.** Scale bars: 500 nm.

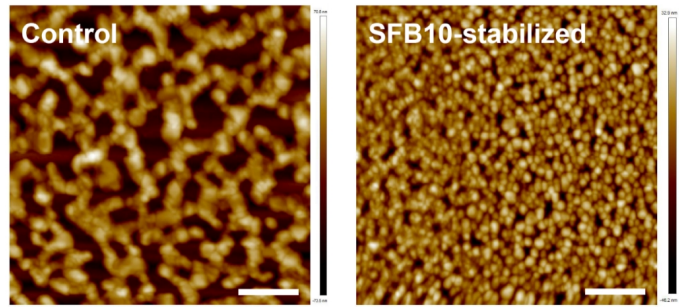

**Extended Data Fig. 6 | AFM images of control and SFB10-stabilized samples.** Scale bars: 1 μm.

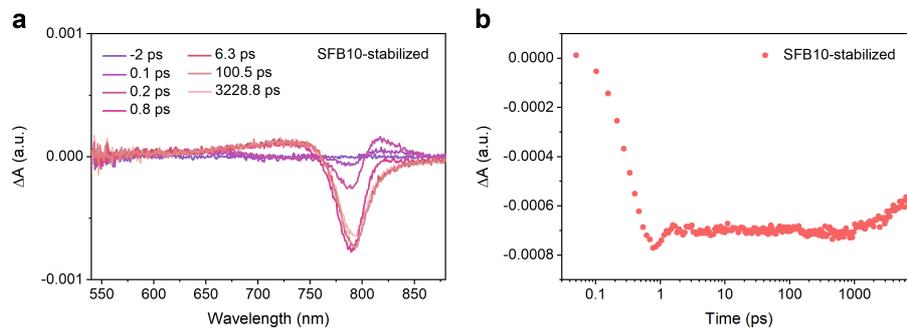

**Extended Data Fig. 7 | Transient absorption (TA) spectra of SFB10-stabilized samples. a**, Transient absorption (TA) spectra of a SFB10-stabilized sample. **b**, The TA dynamics probed at 790 nm.

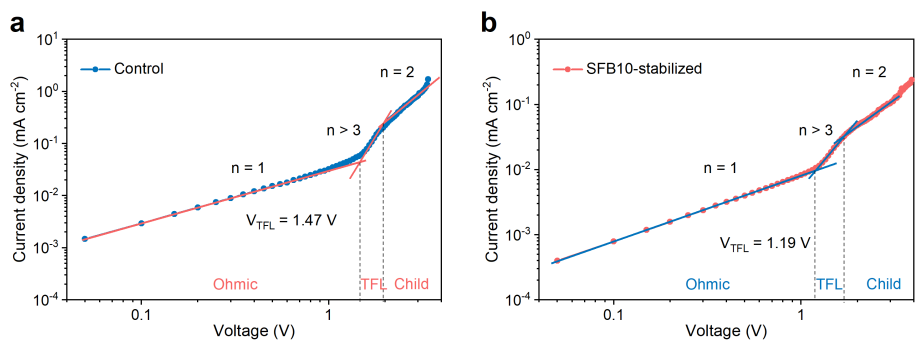

**Extended Data Fig. 8 | Current-voltage characteristics of control and SFB10-stabilized samples using an electron-only device configuration. a**, The current-voltage characteristics of an electron-only device based on the control sample. **b**, The current-voltage characteristics of an electron-only device based on SFB10-stabilized perovskite. The device structure was glass/ITO/ZnO/PEIE/perovskite/TPBi/LiF/Al.

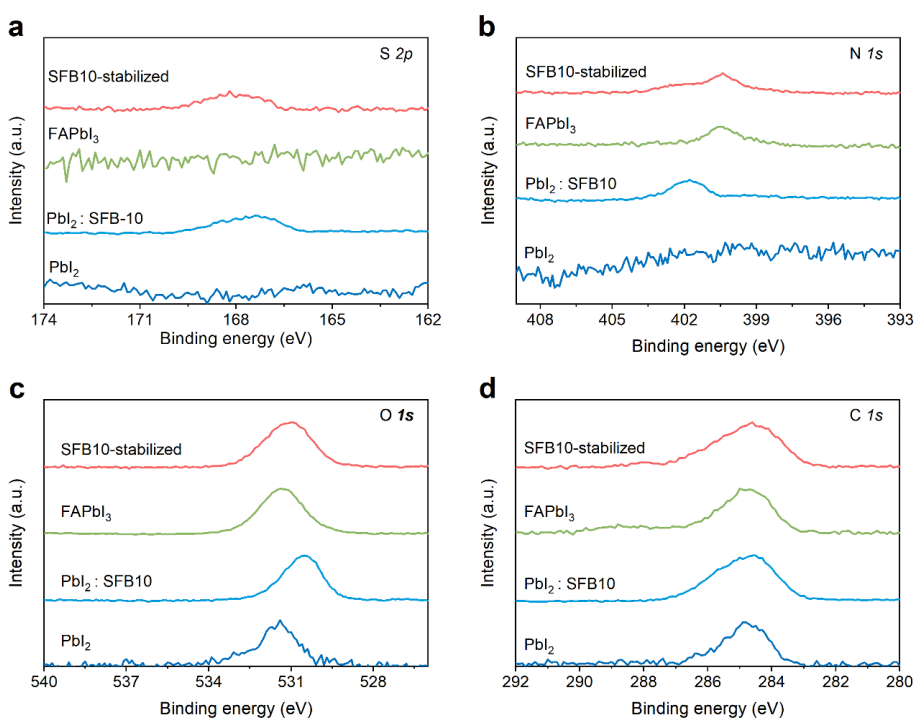

**Extended Data Fig. 9 | XPS measurements of SFB10-stabilized $FAPbI_3$, pristine $FAPbI_3$, SFB10-treated $PbI_2$, and pristine $PbI_2$. a**, S *2p*; **b**, N *1s*; **c**, O *1s*; **d**, C *1s* spectra of samples for SFB10-stabilized $FAPbI_3$, pristine $FAPbI_3$, SFB10-treared $PbI_2$ and pristine $PbI_2$. The XPS spectra were calibrated with C *1s* peak at 284.8 eV.

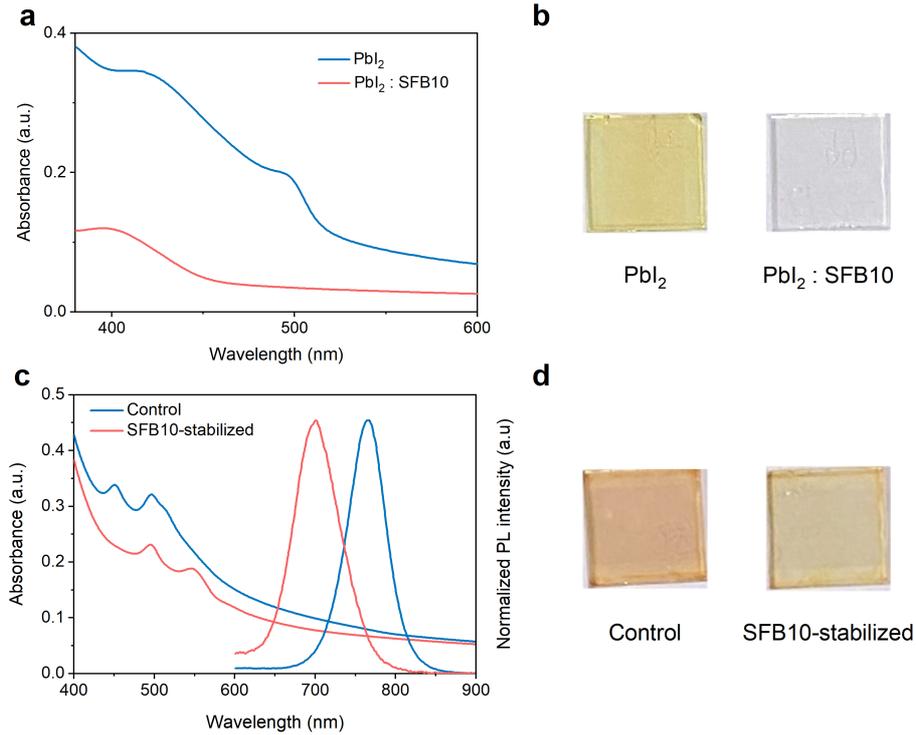

**Extended Data Fig. 10 | Additional optical measurements for SFB10-stabilized FAPbI$_3$, pristine FAPbI$_3$, SFB10-treated PbI$_2$, and pristine PbI$_2$. a,** Absorption spectra of PbI$_2$ and PbI$_2$ SFB10 samples. **b,** Photos of pristine PbI$_2$ and SFB10-treated PbI$_2$ samples after annealing. **c,** Absorption and PL spectra of control and SFB10-stabilized samples before annealing. The molar ratio of SFB10 to PbI$_2$ was 1:1. **d,** Photos of control and SFB10-stabilized perovskite samples before annealing.

**Extended Data Table 1 | Operational lifetime data for SFB10-stabilized PeLEDs.** The average radiance values at each current density are calculated by taking average of ten representative devices. The lifetime estimation for the ongoing tests at lower radiance settings [8.1 W sr$^{-1}$ m$^{-2}$ (10 mA cm$^{-2}$) or below] was based on an analysis of 35 data points from completed T$_{50}$ measurements at 20-200 mA cm$^{-2}$.

| Perovskite composition | Peak EQE (%) | Current density (mA cm$^{-2}$) | EQE@J (champion) | Radiance (W sr$^{-1}$ m$^{-2}$) (average) | T$_{50}$ lifetime (h) (average) |
|---|---|---|---|---|---|
| FAPbI$_3$ (SFB10-stabilized) | 22.8 (@17 mA cm$^{-2}$) | 0.7 | 3.7% | 0.21 | ~1.9×10$^6$ |
| | | 1.1 | 13.6% | 0.42 | ~5.4×10$^5$ |
| | | 3.2 | 18.5% | 2.1 | ~28291 |
| | | 5 | 20.6 % | 3.7 | ~10136 |
| | | 10 | 22.5% | 8.1 | ~2381 |
| | | 20 | 22.7% | 16.5 | 601.2 (max=877.1 h) |
| | | 50 | 21.7% | 38.8 | 131.2 (max=195.3 h) |
| | | 100 | 20.1% | 69.6 | 59.6 (max=120.3 h) |
| | | 200 | 17.9% | 118.8 | 14.1 (max=22.4 h) |